\documentclass{article}

\newenvironment{reminderclaim}[1]{\medskip
\noindent {\bf Reminder of Claim #1.  }\em}{}

\def \QED {\hfill{$\Box$}}

\makeatletter
\newcommand{\eatNext}{\@ifnextchar.\@gobble{}}
\makeatother

\newfont{\mycrnotice}{ptmr8t at 7pt}
\newfont{\myconfname}{ptmri8t at 7pt}
\author{ 
Jeremiah Blocki \\ Carnegie Mellon University\\  \texttt{jblocki@cs.cmu.edu}  \and 
Manuel Blum \\ Carnegie Mellon University\\  \texttt{mblum@cs.cmu.edu}  
\and 
Anupam Datta \\ Carnegie Mellon University\\  \texttt{danupam@cmu.edu}
}

\usepackage{amsmath}

\usepackage{amsfonts, amssymb}
\usepackage{algorithm}
\usepackage[noend]{algpseudocode}
\usepackage{caption}
\usepackage{subcaption}
\usepackage{color}
\usepackage{multirow}
\usepackage{hyperref}

\usepackage{tikz}
\usetikzlibrary{shadows,arrows,decorations,decorations.shapes,backgrounds,shapes}

\newcount\Comments
\definecolor{purple}{rgb}{1,0,1}
\definecolor{darkgreen}{rgb}{0,0.7,0}
\newcommand{\kibitz}[2]{\ifnum\Comments=1\textcolor{#1}{#2}\fi}

\usepackage{graphicx}
\newtheorem{theorem}{Theorem}

\usepackage{multirow}
\usepackage{newfloat}
\newtheorem{claim}{Claim}
\newtheorem{definition}{Definition}
\newcommand{\cut}[1]{}
\newcommand{\paren}[1]{\left(#1\right)}

\newenvironment{proofof}[1]{\noindent {\em Proof of #1.  }}{\QED}

\DeclareFloatingEnvironment[
    fileext=los,
    listname=List of Protocols,
    name=Protocol,
    placement=tbhp,
    within=section,
]{protocol}

\newcommand{\TheTitle}{GOTCHA Password Hackers!}
\newtheorem{assumption}{Assumption}
\begin{document}

%%%%%%%%%%%%%%%%%%%%%%%%%%%%%%%%%%%%%%%%%%%%%%%%%%%%%%%%%%%%%%%%%%%
% title
%%%%%%%%%%%%%%%%%%%%%%%%%%%%%%%%%%%%%%%%%%%%%%%%%%%%%%%%%%%%%%%%%%%
\title{\TheTitle\thanks{This work was partially supported by the NSF Science and Technology TRUST and the AFOSR MURI on Science of Cybersecurity. The first author was also partially supported by an NSF Graduate Fellowship.}} 

%%%%%%%%%%%%%%%%%%%%%%%%%%%%%%%%%%%%%%%%%%%%%%%%%%%%%%%%%%%%%%%%%%%
% authors
%%%%%%%%%%%%%%%%%%%%%%%%%%%%%%%%%%%%%%%%%%%%%%%%%%%%%%%%%%%%%%%%%%%

\maketitle

\begin{abstract}
We introduce GOTCHAs (Generating panOptic Turing Tests to Tell Computers and Humans Apart) as a way of preventing automated offline dictionary attacks against user selected passwords. A GOTCHA is a randomized puzzle generation protocol, which involves interaction between a computer and a human. Informally, a GOTCHA should satisfy two key properties: (1) The puzzles are easy for the human to solve. (2) The puzzles are hard for a computer to solve even if it has the random bits used by the computer to generate the final puzzle --- unlike a CAPTCHA \cite{captcha}. Our main theorem demonstrates that GOTCHAs can be used to mitigate the threat of offline dictionary attacks against passwords by ensuring that a password cracker must receive constant feedback from a human being while mounting an attack. Finally, we provide a candidate construction of GOTCHAs based on Inkblot images. Our construction relies on the usability assumption that users can {\em recognize} the phrases that they originally used to describe each Inkblot image --- a much weaker usability assumption than previous password systems based on Inkblots which required users to recall their phrase exactly. We conduct a user study to evaluate the usability of our GOTCHA construction. We also generate a GOTCHA challenge where we encourage artificial intelligence and security researchers to try to crack several passwords protected with our scheme.   
\end{abstract}

%\category{K.6.5}{Computing Milieux}{Security and Protection}[Authentication]
%A category including the fourth, optional field follows...
%\category{D.2.8}{Software Engineering}{Metrics}[complexity measures, performance measures]

%\terms{ Security, Theory, Inkblots, Artificial Intelligence}
%\keywords{Human Authentication; Passwords; GOTCHA; Inkblots; Offline Dictionary Attack; CAPTCHA; HOSP}

\section{Introduction} \label{sec:Introduction}
Any adversary who has obtained the cryptographic hash of a user's password can mount an automated brute-force attack to crack the password by comparing the cryptographic hash of the user's password with the cryptographic hashes of likely password guesses. This attack is called an offline dictionary attack, and there are many password crackers that an adversary could use \cite{JTR}. Offline dictionary attacks against passwords are  --- unfortunately ---  powerful and commonplace \cite{PasswordCrackingArticle}.  Adversaries have been able to compromise servers at large companies (e.g., Zappos, LinkedIn, Sony, Gawker
\cite{breach:Zappos,breach:sony,breach:militaryHACK,breach:linkedin,breach:rockyou,breach:IEEE}) resulting in the release of millions of cryptographic password hashes \footnote{In a few of these cases \cite{breach:IEEE,breach:rockyou} the passwords were stored in the clear.}. It has been repeatedly demonstrated that users tend to select easily guessable passwords \cite{rockYouPasswords,mostPopularPasswords2012,bonneau2012science}, and password crackers are able to quickly break many of these passwords\cite{seeley1989password}. Offline attacks are becoming increasingly dangerous as computing hardware improves --- a modern GPU can evaluate a cryptographic hash function like SHA2 about 250 million times per second \cite{zonenberg2009distributed} --- and as more and more training data --- leaked passwords from prior breaches --- becomes available \cite{PasswordCrackingArticle}. Symantec reported that compromised passwords have significant economic value to an adversary (e.g., compromised passwords are sold on black market for between \$4 and \$30 ) \cite{passwordBlackMarket}.

HOSPs (Human-Only Solvable Puzzles) were suggested by Canetti, Halevi and Steiner as a way of defending against offline dictionary attacks  \cite{canetti2006mitigating}. The basic idea is to change the authentication protocol so that human interaction is {\em required} to verify a password guess. The authentication protocol begins with the user entering his password. In response the server randomly generates a challenge --- using the password as a source of randomness --- for the user to solve. Finally, the server appends the user\rq{}s response to the user\rq{}s password, and verifies that the hash matches the record on the server. To crack the user\rq{}s password offline the adversary must simultaneously guess the user\rq{}s password and the answer to the corresponding puzzle. The challenge should be easy for a human to solve consistently so that a legitimate user can authenticate. To mitigate the threat of an offline dictionary attack the HOSP should be difficult for a computer to solve --- even if it has all of the random bits used to generate the challenge. 

The basic HOSP construction proposed by Canetti et al. \cite{canetti2006mitigating} was to to fill a hard drive with regular CAPTCHAs (e.g., distorted text) by storing the puzzles without the answers. This solution only provides limited protection against an adversary because the number of unique puzzles that can be generated is bounded by the size of the hard drive (e.g., the adversary could pay people to solve all of the puzzles on the hard drive). See appendix \ref{apdx:Economics} for more discussion. Finding a usable HOSP construction which does not rely on a very large dataset of pregenerated CAPTCHAs is an open problem. Several candidate HOSPs were experimentally tested \cite{poshExperiment} (they are called POSHs in the second paper), but the usability results were underwhelming. 

\paragraph{Contributions} We introduce a simple modification of HOSPs that we call GOTCHAs (Generating panOptic Turing Tests to Tell Computers and Humans Apart). We use the adjective Panoptic to refer to a world without privacy --- there are no hidden random inputs to the puzzle generation protocol. The basic goal of GOTCHAs is similar to the goal of HOSPs --- defending against offline dictionary attacks. GOTCHAs differ from HOSPs in two ways (1) Unlike a HOSP a GOTCHA may require human interaction during the {\em generation} of the challenge.  (2) We relax the requirement that a user needs to be able to answer all challenges easily and consistently. If the user can remember his password during the authentication protocol then he will only ever see one challenge. We only require that the user must be able to answer this challenge consistently. If the user enters the wrong password during authentication then he may see new challenges. We do not require that the user must be able to solve these challenges consistently because authentication will fail in either case. We do require that it is difficult for a computer to distinguish between the ``correct\rq\rq{} challenge and an ``incorrect\rq\rq{} challenge.  Our main theorem demonstrates that GOTCHAs like HOSPs can be used to defend against offline dictionary attacks. The goal of these relaxations is to enable the design of usable GOTCHAs.
 
\begin{figure}
\centering
\includegraphics[width=0.7 \textwidth]{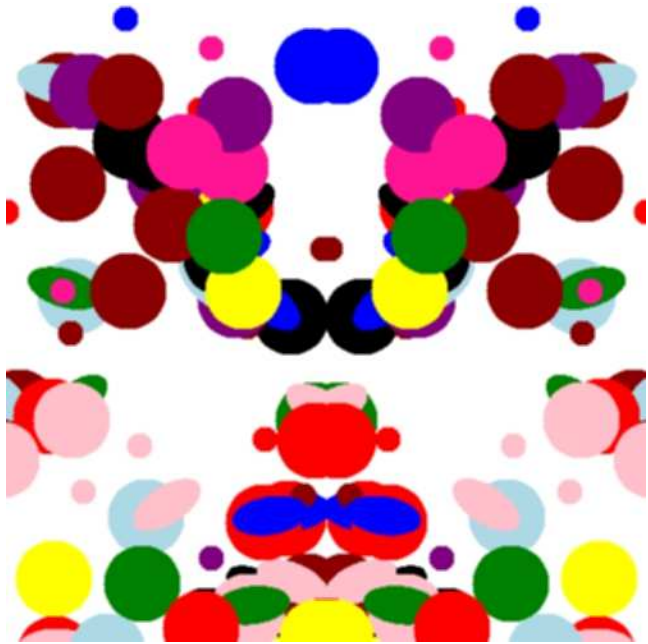}
\caption{Randomly Generated Inkblot Image---An evil clown?} \label{fig:inkblot}
\end{figure}

We introduce a candidate GOTCHA construction based on Inkblot images. While the images are generated randomly by a computer, the human mind can easily imagine semantically meaningful objects in each image. To generate a challenge the computer first generates ten inkblot images (e.g., figure \ref{fig:inkblot}). The user then provides labels for each image (e.g., evil clown, big frog). During authentication the challenge is to match each inkblot image with the corresponding label. We empirically evaluate the usability of our inkblot matching GOTCHA construction by conducting a user study on Amazon\rq{}s Mechanical Turk. Finally, we challenge the AI community to break our GOTCHA construction.

\paragraph{Organization} The rest of the paper is organized as follows: We next discuss related work in section \ref{subsec:related}. We formally define GOTCHAs in section \ref{sec:definitions} and formalize the properties that a GOTCHA should satisfy. We present our candidate GOTCHA construction in section \ref{sec:InkblotConstruction}, and in section \ref{subsec:Authentication} we demonstrate how our GOTCHA could be integrated into an authentication protocol.  We present the results from our user study in section \ref{subsec:UserStudy}, and in section \ref{subsec:OpenChallenge} we challenge the AI and security communities to break our GOTCHA construction. In section \ref{sec:analysis} we prove that GOTCHAs like HOSPs can also be used to design a password storage system which mitigates the threat of offline attacks. We conclude by discussing future directions and challenges in section \ref{sec:disc}.

\subsection{Related Work} \label{subsec:related}

Inkblots \cite{stubblefield2004inkblot} have been proposed as an alternative way to generate and remember passwords. Stubblefield and Simon proposed showing the user ten randomly generated inkblot images, and having the user make up a word or a phrase to describe each image. These phrases were then used to build a 20 character password (e.g., users were instructed to take the first and last letter of each phrase). Usability results were moderately good, but users sometimes had trouble remembering their association. Because the Inkblots are publicly available there is also a security concern that Inkblot passwords could be guessable if different users consistently picked similar phrases to describe the same Inkblot. 

We stress that our use of Inkblot images is different in two ways: (1) Usability: We do not require users to recall the word or phrase associated with each Inkblot. Instead we require user\rq{}s to recognize the word or phrase associated with each Inkblot so that they can match each phrase with the appropriate Inkblot image. Recognition is widely accepted to be easier than the task of recall \cite{baddeley1997,watkins1979appreciation}.  (2) Security: We do not need to assume that it would be difficult for other humans to match the phrases with each Inkblot. We only assume that it is difficult for a computer to perform this matching automatically. 

CAPTCHAs --- formally introduced by Von Ahn et al. \cite{captcha} --- have gained widespread adoption on the internet to prevent bots from automatically registering for accounts. A CAPTCHA is a program that generates a puzzle --- which should be easy for a human to solve and difficult for a computer to solve --- as well as a solution. Many popular forms of CAPTCHAs (e.g., reCAPTCHA \cite{recaptcha}) generate garbled text, which is easy \footnote{Admitedly some people would dispute the use of the label `easy.'} for a human to read, but difficult for a computer to decipher. Other versions of CAPTCHAs rely on the natural human capacity for audio \cite{sauer2008towards} or image recognition \cite{elson2007asirra}. 

CAPTCHAs have been used to defend against online password guessing attacks --- users are sometimes required to solve a CAPTCHA before signing into their account. An alternative approach is to lock out a user after several incorrect guesses, but this can lead to denial of service attacks \cite{dailey2004text}. However, if the adversary has access to the cryptographic hash of the user's password, then he can circumvent all of these requirements and execute an automatic dictionary attack to crack the password offline.  By contrast HOSPs --- proposed by Canetti et al.\cite{canetti2006mitigating} --- were proposed to defend against offline attacks. HOSPs are in some ways similar to CAPTCHAs (Completely Automated Turing Tests to Tell Computers and Humans Apart) \cite{captcha}. CAPTCHAs are widely used on the internet to fight spam by preventing bots from automatically registering for accounts. In this setting a CAPTCHA is sent to the user as a challenge, while the secret solution is used to grade the user's answer. The implicit assumption is that the answer and the random bits used to generate the puzzle remain hidden --- otherwise a spam bot could simply regenerate the puzzle and the answer. While this assumption may be reasonable in the spam bot setting, it does not hold in our offline password attack setting in which the server has already been breached.  A HOSP is different from a CAPTCHA in several key ways: (1) The challenge must remain difficult for a computer to solve even if the random bits used to generate the puzzle are made public. (2) There is no single correct answer to a HOSP. It is okay if different people give different responses to a challenge as long as people can respond to the challenges easily, and each user can consistently answer the challenges.

The only HOSP construction proposed in \cite{canetti2006mitigating} involved stuffing a hard drive with unsolved CAPTCHAs. The problem of finding a HOSP construction that does not rely on a dataset of unsolved CAPTCHAs was left as an open problem \cite{canetti2006mitigating}. Several other candidate HOSP constructions have been experimentally evaluated in subsequent work \cite{poshExperiment} (they are called POSHs in the second paper), but the usability results for every scheme that did not rely on a large dataset on unsolved CAPTCHAs were underwhelming.

 GOTCHAs are very similar to HOSPs. The basic application --- defending against offline dictionary attacks --- is the same as are the key tools: exploiting the power of interaction during authentication, exploiting hard artificial intelligence problems. While the authentication with HOSPs is interactive, the initial generation of the puzzle is not. By contrast, our GOTCHA construction requires human interaction during the initial generation of the puzzle. This simple relaxation allows for the construction of new solutions. In the HOSP paper humans are simply modeled as a puzzle solving oracle, and the adversary is assumed to have a limited number of queries to a human oracle. We introduce a more intricate model of the human agent with the goal of designing more usable constructions.

\paragraph{Password Storage} Password storage is an incredibly challenging problem. Adversaries have been able to compromise servers at many large companies (e.g., Zappos, LinkedIn, Sony, Gawker
\cite{breach:Zappos,breach:sony,breach:militaryHACK,breach:linkedin,breach:rockyou,breach:IEEE}). For example, hackers were able to obtain 32 million plaintext passwords from RockYou using a simple SQL injection attack \cite{breach:rockyou}. While it is considered an extremely poor security practice to store passwords in the clear \cite{noPlaintextPassword}, the practice is still fairly common \cite{bonneau2010password,breach:IEEE,breach:rockyou}. Many other companies \cite{breach:linkedin,bonneau2010password} have used cryptographic hashes to store their passwords, but failed to adopt the practice of salting (e.g., instead of storing the cryptographic hash of the password $h(pw)$ the server stores $\paren{h\paren{pw,r},r}$ for a random string $r$ \cite{salt}) to defend against rainbow table attacks. Rainbow tables, which consist of precomputed hashes, are often used by an adversary to significantly speed up a password cracking attack because the same table can be reused to attack each user when the passwords are unsalted \cite{rainbowTable}. 

Cryptographic hash functions like SHA1, SHA2 and MD5 --- designed for fast hardware computation --- are popular choices for password hashing. Unfortunately, this allows an adversary to try up to 250 million guesses per second on a modern GPU \cite{zonenberg2009distributed}. The BCRYPT \cite{bcrypt} hash function was designed specifically with passwords in mind --- BCRYPT was intentionally designed to be slow to compute (e.g., to limit the power of an  adversary's offline attack). The BCRYPT hash function takes a parameter which allows the programmer to specify how costly the hash computation should be. The downside to this approach is that it also increases costs for the company that stores the passwords (e.g., if we want it to cost the adversary \$1,000 for every million guesses then it will also cost the company at least \$1,000 for every million login attempts). 

Users are often advised (or required) to follow strict guidelines when selecting their password (e.g., use a mix of upper/lower case letters, include numbers and change the password frequently) \cite{NIST-Passwords}. However, empirical studies show that user's are are often frustrated by restricting policies and commonly forget their passwords \cite{komanduri2011passwords,kruger2008empirical,florencio2007large} \footnote{In fact the resulting passwords are sometimes more vulnerable to an offline attack! \cite{komanduri2011passwords,kruger2008empirical}}. Furthermore, the cost of these restrictive policies can be quite high. For example, a Gartner case study \cite{costOfPasswordReset} estimated that it cost over \$17 per password-reset call. Florencio and Herley~\cite{florencio2010security} studied the economic factors that institutions consider before adopting password policies and found that they often value usability over security.

\section{Definitions} \label{sec:definitions}
In this section we seek to establish a theoretical basis for GOTCHAs. Several of the ideas behind our definitions are borrowed from theoretical definitions of CAPTCHAs \cite{captcha} and HOSPs \cite{canetti2006mitigating}. Like CAPTCHAs and HOSPs, GOTCHAs are based on the assumption that some AI problem is hard for a computer to solve, but easy for a person to solve. Ultimately, these assumptions are almost certainly false (e.g., because the human brain can solve a GOTCHA it is reasonable to believe that there exists a computer program to solve the problems). However, it may still be reasonable to assume that these problems cannot be solved by applying {\em known} ideas. By providing a formal definition of GOTCHAs we can determine whether or not a new {\em idea} can be used to break a candidate GOTCHA construction.

We use $c \in \mathcal{C}$ to denote the space of challenges that might be generated. We use $\mathcal{H}$ to denote the set of human users and $H\paren{c, \sigma_t}$ to denote the response that a human $H \in \mathcal{H}$ gives to the challenge $c \in \mathcal{C}$ at time $t$. Here, $\sigma_t$ denotes the state of the human\rq{}s brain at time $t$.  $\sigma_t$ is supposed to encode our user\rq{}s existing knowledge (e.g., vocabulary, experiences) as well as the user\rq{}s mental state at time $t$ (e.g., what is the user thinking about at time $t$). Because $\sigma_t$ changes over time (e.g., new experiences) we use $H\paren{c} = \left\{H\paren{c,\sigma_t}~\vline~t \in \mathbb{N}\right\}$ to denote the {\em set} of all answers a human might give to a challenge $c$. We use $\mathcal{A}$ to denote the range of possible responses (answers) that a human might give to the challenges.

\begin{definition}
Given a metric $d:\mathcal{A}\times\mathcal{A}\rightarrow \mathbb{R}$, we say that a human $H$ can {\em consistently solve} a challenge $c \in \mathcal{C}$ with accuracy $\alpha$ if $\forall t \in \mathbb{N}$
 \[ d\paren{H\paren{c,\sigma_0},H\paren{c,\sigma_t}} \leq \alpha \ , \]
where $\sigma_0$ denotes the state of the human's brain when he initially answers the challenge. If $\left|H\paren{c}\right|=1$ then we simply say that the human can consistently solve the challenge. 
\end{definition} 

{\bf \noindent Notation:} When we have a group of challenges $\langle c_1,\ldots,c_k \rangle$ we will sometimes write $H\paren{\langle c_1,\ldots,c_k \rangle,\sigma_t} =$ \\$\langle H\paren{c_1,\sigma_t},\ldots,H\paren{c_k,\sigma_t}\rangle$ for notational convenience. We use $y \sim \mathcal{D}$ to denote a random sample from the distribution $\mathcal{D}$, and we use $r \stackrel{\$}{\gets} \{0,1\}^n$ to denote a element drawn from the set $\{0,1\}^n$ uniformly at random.

One of the requirements of a HOSP puzzle system \cite{canetti2006mitigating} is that the human $H$ must be able to {\em consistently} answer {\em any} challenge that is generated (e.g., $\forall c \in \mathcal{C},$ $H$ can consistently solve $c$). These requirements seem to rule out promising ideas for HOSP constructions like Inkblots\cite{poshExperiment}. In this construction the challenge is a randomly generated inkblot image $I$, and the response $H\paren{I, \sigma_0}$ is word or phrase describing what the user initially sees in the inkblot image (e.g., evil clown, soldier, big lady with a ponytail). User studies have shown that $H\paren{I, \sigma_0}$ does not always match $H\paren{I, \sigma_t}$ --- the phrase describing what the user sees at time $t$ \cite{poshExperiment}. In a few cases the errors may be correctable (e.g., capitalization, plural/singular form of a word), but oftentimes the phrase was completely different  --- especially if a long time passed in between trials \footnote{We would add the requirement that the human must be able to consistently answer the challenges without spending time memorizing and rehearsing his response to the challenge. Otherwise we could just as easily force the user to remember a random string to append on to his password.}.  By contrast, our GOTCHA construction does not require the user to remember the phrases associated with each Inkblot. Instead we rely on a much weaker assumption --- the user can consistently recognize his solutions. We say that a human can recognize his solutions to a set of challenges if he can consistently solve a matching challenge (definition \ref{assum:UsabilityMatching}) in which he is asked to match each of his solutions with the corresponding challenge.

\begin{definition} \label{def:MatchingChallenge} \label{assum:UsabilityMatching}
Given an integer $k$, and a permutation $\pi:[k]\rightarrow[k]$, a {\em matching challenge} $\hat{c}_{\pi} = \paren{\vec{c},\vec{a}} \in \mathcal{C}$ of size $k$ is given by a k-tuple of challenges $\vec{c} = \langle c_{\pi\paren{1}},\ldots,c_{\pi(k)} \rangle \in \mathcal{C}^k$ and solutions $\vec{a} = H\paren{\langle c_1,\ldots, c_k \rangle,\sigma_0}$. The response to a matching challenge is a permutation $\pi' = H\paren{\vec{c}_{\pi},\sigma_t}$.
\end{definition}

For permutations $\pi:[k]\rightarrow[k]$ we use the distance metric 
\[ d_k\paren{\pi_1,\pi_2} = \left|\left\{i ~\vline~\pi_1(i) \neq \pi_2(i) \wedge 1 \leq i \leq k \right\} \right| \ . \]
$d_k\paren{\pi_1,\pi_2}$ simply counts the number of entries where the permutations don't match. We say that a human can consistently recognize his solution to a matching challenge  $\hat{c}_{\pi}$ with accuracy $\alpha$ if $\forall  t. d_k\paren{H\paren{\hat{c}_{\pi},\sigma_t},\pi} \leq \alpha$. We use $\{\pi'~\vline~d_k\paren{\pi,\pi'} \leq \alpha \}$ to denote the set of permutations $\pi'$ that are $\alpha$-close to $\pi$.

The puzzle generation process for a GOTCHA involves interaction between the human and a computer: (1) The computer generates a set of $k$ challenges. (2) The human solves these challenges. (3) The computer uses the solutions to produce a final challenge \footnote{We note that a HOSP puzzle system $\paren{G}$ \cite{canetti2006mitigating} can be modeled as a GOTCHA puzzle system $\paren{G_1,G_2}$ where $G_1$ does nothing and $G_2$ simply runs $G$ to generate the final challenge $\hat{c}$ directly.}. Formally, 

\begin{definition}\label{def:PuzzleSystem} A puzzle-system is a pair $\paren{G_1,G_2}$, where $G_1$ is a randomized challenge generator that takes as input $1^k$ (with $k$ security parameter) and a pair of random bit strings $r_1,r_2 \in \{0,1\}^*$ and outputs $k$ challenges $\langle c_1,\ldots ,c_k\rangle \leftarrow G_1\paren{1^k,r_1,r_2}$. $G_2$ is a randomized challenge generator that takes as input $1^k$ (security parameter), a random bit string $r_1 \in \{0,1\}^*$, and proposed answers $\vec{a}=\langle a_1,...,a_k\rangle$ to the challenges $G_1\paren{1^k,r_1,r_2}$ and outputs a challenge\\ $\hat{c} \gets G_2\paren{1^k,r_1,\vec{a}}$. We say that the puzzle-system is $\paren{\alpha,\beta}$-{\em usable} if \[\Pr_{H \stackrel{\$}{\gets} \mathcal{H}}\left[ \mathbf{Accurate}\paren{H,\hat{c},\alpha} \right] \geq \beta \ , \]
whenever $\vec{a}  = H\paren{G_1\paren{1^k,r_1,r_2},\sigma_0}$, where $\mathbf{Accurate}\paren{H,\hat{c},\alpha}$ denotes the event that
 the human  $H$ can {consistently} solve $\hat{c}$ with accuracy $\alpha$.
\end{definition}

In our authentication setting the random string $r_1$ is extracted from the user's password using a strong pseudorandom function $\mathbf{Extract}$. To provide a concrete example of a puzzle-system, $G_1$ could be a program that  generates a set of inkblot challenges $\langle I_1,\ldots,I_k\rangle $ using random bits $r_1$, selects a random permutation $\pi:[k]\rightarrow [k]$ using random bits $r_2$, and returns $\langle I_{\pi(1)},\ldots,I_{\pi(k)}\rangle$. The human's response to an Inkblot --- $H\paren{I_j,\sigma_0}$ --- is whatever he/she imagines when he sees the inkblot $I_j$ for the first time   (e.g., some people might imagine an evil clown when they look at figure \ref{fig:inkblot}). Finally, $G_2$ might  generate Inkblots $\vec{c}=\langle I_1,\ldots,I_k\rangle $ using random bits $r_1$, and return the matching challenge $\hat{c}_\pi = \paren{\vec{c},\vec{a}}$. In this case the matching challenge is for the user to match his labels with the appropriate Inkblot images to recover the permutation $\pi$. Observe that the final challenge --- $\hat{c}_\pi$ --- can only be generated after a round of {\em interaction} between the computer and a human. By contrast, the challenges in a HOSP must be generated automatically by a computer. Also notice that if $G_2$ is executed with a different random bit string $r_1'$ then we do not require the resulting challenge to be consistently recognizable (e.g., if the user enters in the wrong password then authentication will fail regardless of how he solves the resulting challenge).  For example, if the user enters the wrong password the user might be asked to match his labels $\langle \ell_{\pi(1)},...,\ell_{\pi(k)}\rangle = H\paren{\langle I_{\pi(1)},\ldots,I_{\pi(k)}\rangle,\sigma_0}$ with Inkblots $\langle I_1',\ldots,I_k'\rangle$ that he has never seen. 

An adversary could attack a puzzle system by either (1) attempting to distinguish between the correct puzzle, and puzzles that might be meaningless to the human, or (2) by solving the matching challenge directly. 

We say that an algorithm $A$ can distinguish distributions $\mathcal{D}_1$ and $\mathcal{D}_2$ with advantage $\epsilon$ if
\[ \left| \Pr_{x \sim \mathcal{D}_1}\left[ A\paren{x} = 1 \right] -  \Pr_{y \sim \mathcal{D}_2}\left[ A\paren{y} = 1 \right] \right| \geq \epsilon \ . \]

Our formal definition of a GOTCHA is found in definition \ref{def:GOTCHA}. Intuitively, definition \ref{def:GOTCHA} says that (1) The underlying puzzle-system should be usable --- so that legitimate users can authenticate. (2) It should be difficult for the adversary to distinguish between the correct matching challenge (e.g., the one that the user will see when he types in the correct password), and an incorrect matching challenge (e.g., if the user enters the wrong password he will be asked to match his labels with different Inkblot images), and (3) It should be difficult for the adversary to distinguish between the user's matching, and a random matching drawn from a distribution $R$ with sufficiently high minimum entropy. 

\begin{definition} \label{def:GOTCHA}
A puzzle-system $\paren{G_1,G_2}$ is an $(\alpha,\beta,\epsilon,\delta, \mu)$-GOTCHA if (1)  $\paren{G_1,G_2}$ is $\paren{\alpha,\beta}$-{\em usable}  (2) Given a human $H \in \mathcal{H}$ no probabilistic polynomial time algorithm can distinguish between distributions \[\mathcal{D}_1 = \left\{  \mathrel{\substack{ H\paren{G_1\paren{1^k,r_1,r_2},\sigma_0}, \\G_2\paren{1^k,r_1,H\paren{G_1\paren{1^k,r_1,r_2},\sigma_0} }}}~\vline~r_1,r_2 \stackrel{\$}{\gets} \{0,1\}^n \right\} \ \] and \[\mathcal{D}_2 = \left\{ \mathrel{\substack{H\paren{G_1\paren{1^k,r_1,r_2},\sigma_0},\\ G_2\paren{1^k,r_3,H\paren{G_1\paren{1^k,r_1,r_2},\sigma_0}}}}~\vline~r_1,r_2,r_3 \stackrel{\$}{\gets} \{0,1\}^n \right\}\] with advantage greater than $\epsilon$, and (3) Given a human $H \in \mathcal{H}$, there is a distribution $R(c)$ with $\mu(m)$ bits of minimum entropy such that no probabilistic polynomial time algorithm can distinguish between distributions  \[\mathcal{D}_3 = \left\{
 \mathrel{\substack{H\paren{G_1\paren{1^k,r_1,r_2},\sigma_0} \\ G_2\paren{1^k,r_1,H\paren{G_1\paren{1^k,r_1,r_2},\sigma_0} }, \\ H\paren{ G_2\paren{1^k,r_1,H\paren{G_1\paren{1^k,r_1,r_2},\sigma_0} },\sigma_0}}}
~\vline~r_1,r_2 \stackrel{\$}{\gets} \{0,1\}^n \right\} \]  and \[\mathcal{D}_4 = \left\{\mathrel{\substack{H\paren{G_1\paren{1^k,r_1,r_2},\sigma_0} \\ G_2\paren{1^k,r_1,H\paren{G_1\paren{1^k,r_1,r_2},\sigma_0}},\\ R\paren{ G_2\paren{1^m,r_1,\langle a_1,...,a_m\rangle },\sigma_0}}}~\vline~r_1,r_2 \stackrel{\$}{\gets} \{0,1\}^n \right\}\] with advantage greater then $\delta$.
\end{definition}

\subsection{Password Storage and Offline Attacks}
To protect users in the event of a server breach organizations are advised to store salted password hashes --- using a cryptographic hash function ($h:\{0,1\}^*\rightarrow \{0,1\}^n$) and a random bit string $(s \in \{0,1\}^*)$ \cite{NIST-Passwords}. For example, if a user (u) chose the password (pw) the server would store the tuple $\paren{u,s,h\paren{s,pw}}$. Any adversary who has obtained  $\paren{u,s,h\paren{s,pw}}$ (e.g., through a server breach) may mount a --- fully automated --- offline dictionary attack using powerful password crackers like John the Ripper~\cite{JTR}. To verify a guess $pw'$ the adversary simply computes $h\paren{s,pw'}$ and checks to see if this hash matches $h\paren{s,pw}$. 

We assume that an adversary $\mathbf{Adv}$ who breaches the server can obtain the code for $h$, as well as the code for any GOTCHAs used in the authentication protocol. Given the code for $h$ and the salt value $s$ the adversary can construct a function 
\[ \mathbf{VerifyHash}\paren{pw'}  = \left\{
\begin{aligned} 
1 && \mbox{if $h\paren{s,pw} = h\paren{s,pw'}$} \\
0 && \mbox{otherwise.}
\end{aligned} \right. \ . \]

We also allow the adversary to have black box access to a GOTCHA solver (e.g., a human). We use $c_H$ to denote the cost of querying a human and $c_h$ to denote the cost of querying the function $\mathbf{VerifyHash}$\footnote{The value of $c_h$ may vary widely depending on the particular cryptographic hash function --- it is inexpensive to evaluate SHA1, but BCRYPT \cite{bcrypt} may be very expensive to evaluate.}, and we use $n_H$ (resp. $n_h$) to denote the number of queries to the human (resp. $\mathbf{VerifyHash}$). Queries to the human GOTCHA solver are much more expensive than queries to the cryptographic hash function ($c_H \gg c_h$) \cite{motoyama2010re}. For technical reasons we limit our analysis to conservative adversaries.
\begin{definition} \label{def:ConservativeAdversary}
We say that an adversary $\mathbf{Adv}$  is {\em conservative} if (1) $\mathbf{Adv}$  uses the cryptographic hash function $h$ in a black box manner (e.g., the hash function $h$ and the stored hash value are only used to construct a subroutine $\mathbf{VerifyHash}$ which is then used as a black box by $\mathbf{Adv}$ ), (2) The pseudorandom function $\mathbf{Extract}$ is used as a black box, and (3) The adversary only queries a human about challenges generated using a password guess.
\end{definition}
It is reasonable to believe that our adversary is conservative. All existing password crackers (e.g., \cite{JTR}) use the hash function as a black box, and it is difficult to imagine that the adversary would benefit by querying a human solver about Inkblots that are unrelated to the password. 

We use $D \subseteq \{0,1\}^*$ to denote a dictionary of likely guesses that the adversary would like to try,  \[\mathbf{Cost}\paren{\mathbf{Adv}, D} = \paren{n_hc_h+n_Hc_H}\] to denote the cost of the queries that the adversary makes to check each guess in $D$, and $\mathbf{Succeed}\paren{\mathbf{Adv}, D, pw}$ to denote the event that the adversary makes a query to  $\mathbf{VerifyHash}$ that returns $1$ (e.g., the adversary successfully finds the user's password $pw$). The adversary might use a computer program to try to solve some of the GOTCHAs --- to save cost by not querying a human. However, in this case the adversary might fail to crack the password because the GOTCHA solver found the wrong solution to one of the challenges.

\begin{definition} \label{def:PasswordCracker}
An adversary $\mathbf{Adv}$ is  $\paren{C, \gamma, D}$-successful if $\mathbf{Cost}\paren{\mathbf{Adv}, D} \leq C$, and \[\Pr_{pw \stackrel{\$}{\gets} D }\left[\mathbf{Succeed}\paren{\mathbf{Adv}, D, pw}\right] \geq \gamma \ .\]
 
\end{definition}

Our attack model is slightly different from the attack model in \cite{canetti2006mitigating}. They assume that the adversary may ask a limited number of queries to a human challenge solution oracle. Instead we adopt an economic model similar to \cite{blockiNaturallyRehearsing}, and assume that the adversary is instead limited by a budget $C$, which may be used to either evaluate the cryptographic hash function $h$ or query a human $H$. 

\section{Inkblot Construction} \label{sec:InkblotConstruction}
Our candidate GOTCHA construction is based on Inkblots images. We use algorithm \ref{alg:GenerateInkblot1} to generate inkblot images. Algorithm \ref{alg:GenerateInkblot1} takes as input random bits $r_1$ and a security parameter $k$ --- which specifies the number of Inkblots to output. Algorithm \ref{alg:GenerateInkblot1} makes use of the randomized subroutine $\mathbf{DrawRandomEllipsePairs}\paren{I, t,width,height}$ which draws $t$ pairs of ellipses on the image $I$ with the specified width and height. The first ellipse in each pair is drawn at a random $(x,y)$ coordinate on the left half of the image with a randomly selected color and angle $\alpha$ of rotation, and the second ellipse is mirrored on the right half of the image. Figure \ref{fig:inkblot} is an example of an Inkblot image generated by algorithm \ref{alg:GenerateInkblot1}.

\begin{algorithm}
\caption{$\mathbf{GenerateInkblotImages}$}
\begin{algorithmic}
\State {\bf Input:} Security Parameter $1^k$, Random bit string $r_1 \in \{0,1\}^*$.
\For{$j = 1,\ldots,k$}
\State $I_j \gets$ new Blank Image
\Comment{The following operations only use the random bit string $r_1$ as a source of randomness}
\State $\mathbf{DrawRandomEllipsePairs}\paren{I_j, 150, 60,60}$
\State $\mathbf{DrawRandomEllipsePairs}\paren{I_j, 70, 20,20}$
\State $\mathbf{DrawRandomEllipsePairs}\paren{I_j, 150, 60,20}$
\EndFor
\noindent \Return $\langle I_{1},\ldots, I_{k} \rangle$ \Comment{Inkblot Images}
\end{algorithmic}
\label{alg:GenerateInkblot1}
\end{algorithm}

Our candidate GOTCHA is given by the pair $\paren{G_1,G_2}$ --- algorithms \ref{alg:GenerateInkblot} and \ref{alg:GenerateMatchingChallenge}. $G_1$ runs algorithm \ref{alg:GenerateInkblot1} to generate $k$ Inkblot images, and then returns these images in permuted order --- using a function \\$\mathbf{GenerateRandomPermutation}\paren{k,r}$, which generates a random permutation $\pi:[k]\rightarrow[k]$ using random bits $r$. $G_2$ also runs algorithm \ref{alg:GenerateInkblot1} to generate $k$ Inkblot images, and then outputs a matching challenge. 

\begin{algorithm}
\caption{$G_1$}
\begin{algorithmic}
\State {\bf Input:} Security Parameter $1^k$, Random bit strings $r_1,r_2 \in \{0,1\}^*$.
\State $\langle I_1,\ldots,I_k \rangle \gets \mathbf{GenerateInkblotImages}\paren{k,r_1}$
\State $\pi \gets \mathbf{GenerateRandomPermutation}\paren{k,r_2}$
\State \Return $\langle I_{\pi(1)},\ldots, I_{\pi(k)} \rangle$ 
\end{algorithmic}
\label{alg:GenerateInkblot}
\end{algorithm}

After the Inkblots $\langle I_{\pi(1)},\ldots, I_{\pi(k)} \rangle$  have been generated, the human user is queried to provide labels $\ell_{\pi(1)},\ldots,\ell_{\pi(k)}$ where 
\[ \langle \ell_{\pi(1)},\ldots,\ell_{\pi(k)}\rangle  = H\paren{\langle I_{\pi(1)},\ldots, I_{\pi(k)} \rangle,\sigma_0} \ . \]
In our authentication setting the server would store the labels $\ell_{\pi(1)},\ldots,\ell_{\pi(k)}$ in permuted order. The final challenge --- generated by algorithm \ref{alg:GenerateMatchingChallenge} --- is to match the Inkblot images $I_1,\ldots,I_k$ with the user generated labels $\ell_1,...,\ell_k$ to recover the permutation $\pi$.

\begin{algorithm}
\caption{$\mathbf{GenerateMatchingChallenge}~~~G_2$}
\begin{algorithmic}
\State {\bf Input:} Security Parameter $1^k$, Random bits $r_1 \in \{0,1\}^*$ and labels $\vec{a} = \langle\ell_{\pi(1)},\ldots,\ell_{\pi(k)}\rangle$.
\State $\langle I_1,\ldots,I_k \rangle  \gets \mathbf{GenerateInkblotImages}\paren{1^k,r_1}$
\State \Return $\hat{c}_\pi = \paren{\vec{c},\vec{a}}$ \Comment{Matching Challenge} 
\end{algorithmic}
\label{alg:GenerateMatchingChallenge}
\end{algorithm}

{\bf Observation: } Notice that if the random bits provided as input to $\mathbf{GenerateInkblotImages}$ and \\ $\mathbf{GenerateMatchingChallenge}$  match that the user will see the same Inkblot images in the final matching challenge. However, if the random bits do not match (e.g., because the user typed the wrong password in our authentication protocol) then the user will see different Inkblot images. The labels $\ell_{1},\ldots,\ell_{k}$ will be the same in both cases.

\cut{
\begin{assumption} \label{assum:UsabilityMatching}
The pair $\paren{G_1,G_2}$ is a $\paren{\alpha,\beta,\epsilon,\delta,\mu}$-GOTCHA.
\end{assumption}
}

\subsection{GOTCHA Authentication} \label{subsec:Authentication}
To illustrate how our GOTCHAs can be used to defend against offline attacks we present the following authentication protocols: {\bf Create Account } (protocol \ref{protocol:createpwd}) and {\bf Authenticate} (protocol \ref{protocol:authenticate}). Communication in both protocols should take place over a secure channel. Both protocols involve several rounds of interaction between the user and the server. To create a new account the user sends his username/password to the server, the server responds by generating $k$ Inkblot images $I_1,\ldots,I_k$, and the user provides a response $\langle \ell_1,\ldots,\ell_k \rangle =  H\paren{\langle I_1,\ldots,I_k\rangle,\sigma_0}$ based on his mental state at the time --- the server stores these labels in permuted order $\ell_{\pi(1)},\ldots,\ell_{\pi(k)} $ \footnote{For a general GOTCHA, protocol \ref{protocol:createpwd} would need to have an extra round of communication. The server would send the user the final challenge generated by $G_2$ and the user would respond with $H\paren{G_2\paren{,},\sigma_0}$. Protocol \ref{protocol:createpwd} takes advantage of the fact that $\pi = H\paren{G_2\paren{,},\sigma_0}$ is already known. }. To authenticate later the user will have to match these labels with the corresponding inkblot images to recover the permutation $\pi$. 

In section \ref{subsec:offlineAttack} we argue that the adversary who wishes to mount a cost effective offline attack needs to obtain constant feedback from a human. Following \cite{canetti2006mitigating} we assume that the function $\mathbf{Extract}:\{0,1\}^*\rightarrow\{0,1\}^n$ is a strong randomness extractor, which can be used to extract random strings from the user's password. Recall that  $h:\{0,1\}^*\rightarrow\{0,1\}^*$ denotes a cryptographic hash function.

\begin{protocol}[htb]
\caption{Create Account} 
\begin{algorithmic}
\State {\bf Security Parameters: } $k$, $n$.
\State {\bf (User):} Select username $(u)$ and password $(pw)$ and send $\paren{u,pw}$ to the server.
\State {\bf (Server): } Sends Inkblots $\langle I_1,\ldots, I_k \rangle$ to the user where:
\State~~~~$r' \stackrel{\$}{\gets} \{0,1\}^n$,~$r_1 \gets \mathbf{Extract}\paren{pw,r'}$, ~$r_2 \stackrel{\$}{\gets} \{0,1\}^n$~ and
\State~~~~$\langle I_{1},\ldots, I_k \rangle \gets \mathbf{GenerateInkblotImages}\paren{1^k,r_1}$
\State {\bf (User): } Sends responses $\langle \ell_1,...,\ell_k\rangle$ back to the server where:
\State~~~~$\langle \ell_1,\ldots,\ell_k\rangle \gets H\paren{\langle I_1,\ldots, I_k\rangle,\sigma_0}$.
\State {\bf (Server): } Store the tuple $t$ where $t$ is computed as follows:
~~~~\State Salt: $s \stackrel{\$}{\gets} \{0,1\}^n$
~~~~\State $\pi \gets \mathbf{GenerateRandomPermutation}\paren{k,r_2}$.
~~~~\State $h_{pw} \gets h\paren{u,s,pw,\pi(1),...,\pi(k)}$
~~~~\State $t \gets \paren{u,r',s,h_{pw}, \ell_{\pi(1)},\ldots,\ell_{\pi(k)}  }$
\end{algorithmic}
\label{protocol:createpwd}
\end{protocol}

\begin{protocol}[htb]
\caption{Authenticate} 
\begin{algorithmic}
\State {\bf Security Parameters: } $k$, $n$.
\State {\bf Usability Parameter: } $\alpha$
\State {\bf (User):} Send username $(u)$ and password $(pw')$ --- $pw'$ may or may not be correct.
\State {\bf (Server): } Sends challenge $\hat{c}$ to the user where $\hat{c}$ is computed as follows:
\State~~~~Find $t =\paren{u,r', s,h_{pw}, \ell_{\pi(1)},\ldots,\ell_{\pi(k)}  }$
\State~~~~$r_1' \gets \mathbf{Extract}\paren{pw',r'}$
\State~~~~$\langle I_1',...,I_k'\rangle \gets \mathbf{GenerateInkblotImages}\paren{r_1',k}$
\State~~~~$\hat{c}_\pi \gets \paren{\langle I_1,...,I_k\rangle,\langle\ell_{\pi(1)},\ldots,\ell_{\pi(k)}\rangle}$
\State {\bf (User): } Solves $\hat{c}_\pi$ and sends the answer $\pi' = H\paren{\hat{c},\sigma_t}$.

\State {\bf (Server): }
\ForAll{$\pi_0$ s.t $d_k\paren{\pi_0,\pi'}\leq \alpha$}
~~~~\State $h_{pw,0} \gets h\paren{u,s,pw',\pi_0(1),...,\pi_0(k)}$
~~~~\If{$h_{pw,0} = h_{pw}$} 
~~~~ \State {\bf Authenticate } 
\EndIf
\EndFor 
\State {\bf Deny } 

\end{algorithmic}
\label{protocol:authenticate}
\end{protocol}

Our protocol could be updated to allow the user to reject challenges he found confusing during account creation in protocol \ref{protocol:createpwd}. In this case the server would simply note that the first GOTCHA was confusing and generate a new GOTCHA. Once our user has created an account he can login by following protocol \ref{protocol:authenticate}.

Claim \ref{claim:correct} says that a legitimate user can successfully authenticate if our Inkblot construction satisfies the usability requirements of a GOTCHA. The proof of claim \ref{claim:correct} can be found in appendix \ref{apdx:missingproof}.

\begin{claim}\label{claim:correct}
If $\paren{G_1,G_2}$ is a $\paren{\alpha,\beta,\epsilon,\delta,\mu}$-GOTCHA then at least $\beta$-fraction of humans can successfully authenticate using protocol \ref{protocol:authenticate} after creating an account using protocol \ref{protocol:createpwd}.
\end{claim}

One way to improve usability of our authentication protocol is to increase the neighborhood of acceptably close matchings by increasing $\alpha$. The disadvantage is that the running time for the server in protocol \ref{protocol:authenticate} increases with the size of $\alpha$. Claim \ref{claim:ClosePermuations} bounds the time needed to enumerate over all close permuations. The proof of claim \ref{claim:ClosePermuations} can be found in appendix \ref{apdx:missingproof}.

\begin{claim} \label{claim:ClosePermuations}
For all permutations $\pi:[k]\rightarrow[k]$ and $\alpha \geq 0$ 
\[ \left|\left\{\pi'~\vline~d_k\paren{\pi,\pi'}\leq \alpha \right\} \right| \leq 1 + \sum_{i=2}^\alpha {k \choose i}i! \ .  \]
\end{claim}

For example, if the user matches $k=10$ Inkblots and we want to accept matchings that are off by at most $\alpha=5$ entries then the server would need to enumerate over at most $36,091$ permutations\footnote{A more precise calculation reveals that there are exactly $13,264$ permutations s.t. $d_{10}\paren{\pi',\pi}\leq 5$ and a random permuation $\pi'$ would only be accepted with probability $3.66\times 10^{-3}$}. Organizations are already advised to use password hash functions like BCRYPT \cite{bcrypt} which intentionally designed to be slower than standard cryptographic hash functions --- often by a factor of millions. Instead of making the hash function a million times slower to evaluate the server might instead make the hash function a thousand times slower to evaluate and use these extra computation cycles to enumerate over close permutations. The organization's trade-off is between: security, usability and the resources that it needs to invest during the authentication process.

We observe that an adversary mounting an online attack would be naturally rate limited because he would need to solve a GOTCHA for each new guess. Protocol \ref{protocol:authenticate} could also be supplemented with a $k$-strikes policy --- in which a user is locked out for several hours after $k$ incorrect login attempts --- if desired.

\subsection{User Study} \label{subsec:UserStudy}
To test our candidate GOTCHA construction we conducted an online user study\footnote{Our study protocol was approved for exemption by the Institutional Review Board (IRB) at Carnegie Mellon University (IRB Protocol Number: HS13-219).}. We recruited participants through Amazon's Mechanical Turk to participate in our study. The study was conducted in two phases. In phase 1 we generated ten random Inkblot images for each participant, and asked each participant to provide labels for their Inkblot images. Participants were advised to use creative titles (e.g.,  evil clown, frog, lady with poofy dress) because they would not need to remember the exact titles that they used. Participants were paid $\$1$ for completing this first phase. A total of $70$ users completed phase 1. 

%See figure \ref{fig:userStudyphase2} for an example of phase 1.

\begin{figure}[h]
\centering
\includegraphics[width=0.7\textwidth]{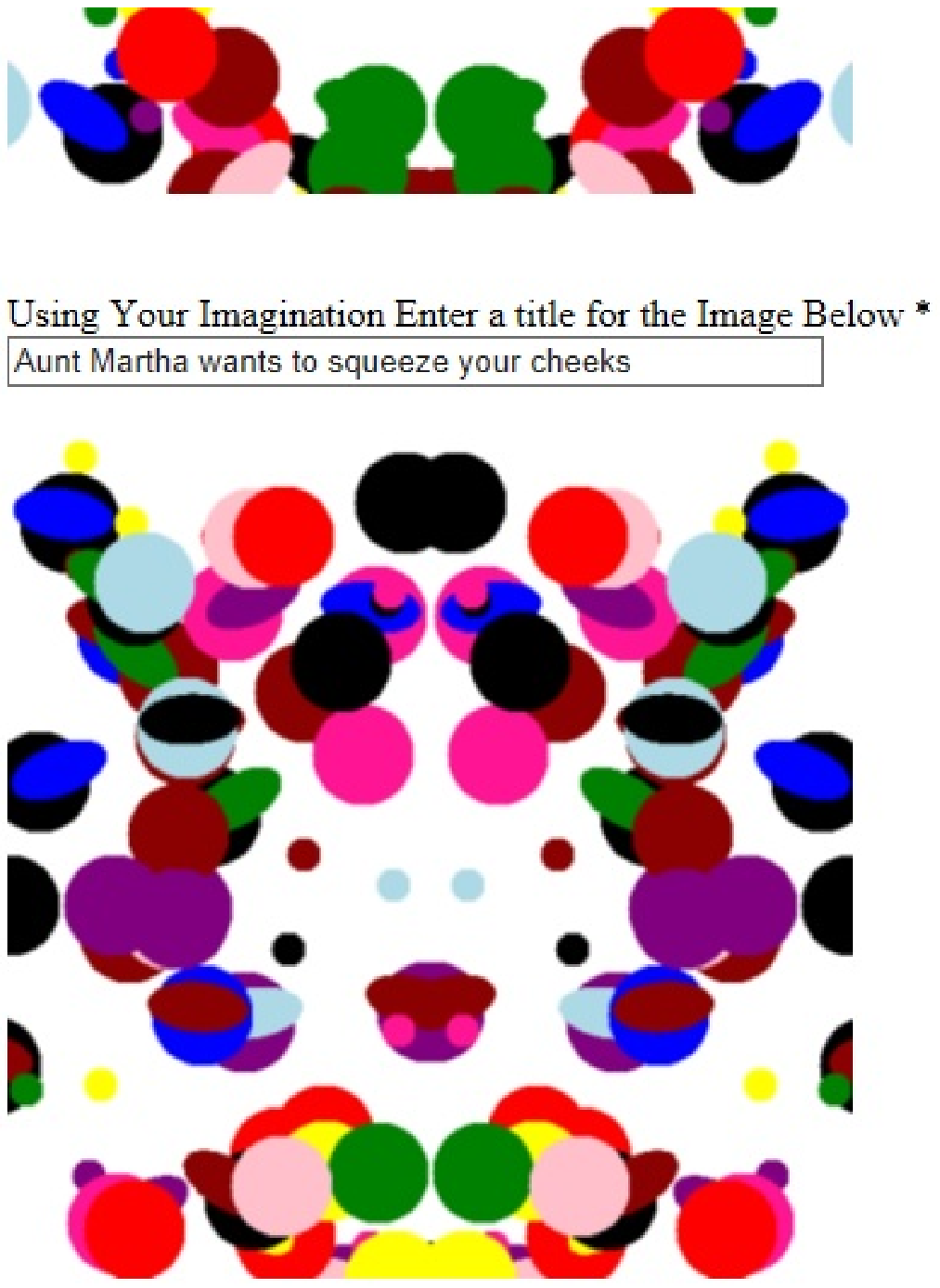}
\caption{Phase 1} \label{fig:userStudyphase1}
\end{figure}

\begin{figure}[h]
\centering
\includegraphics[width=0.7\textwidth]{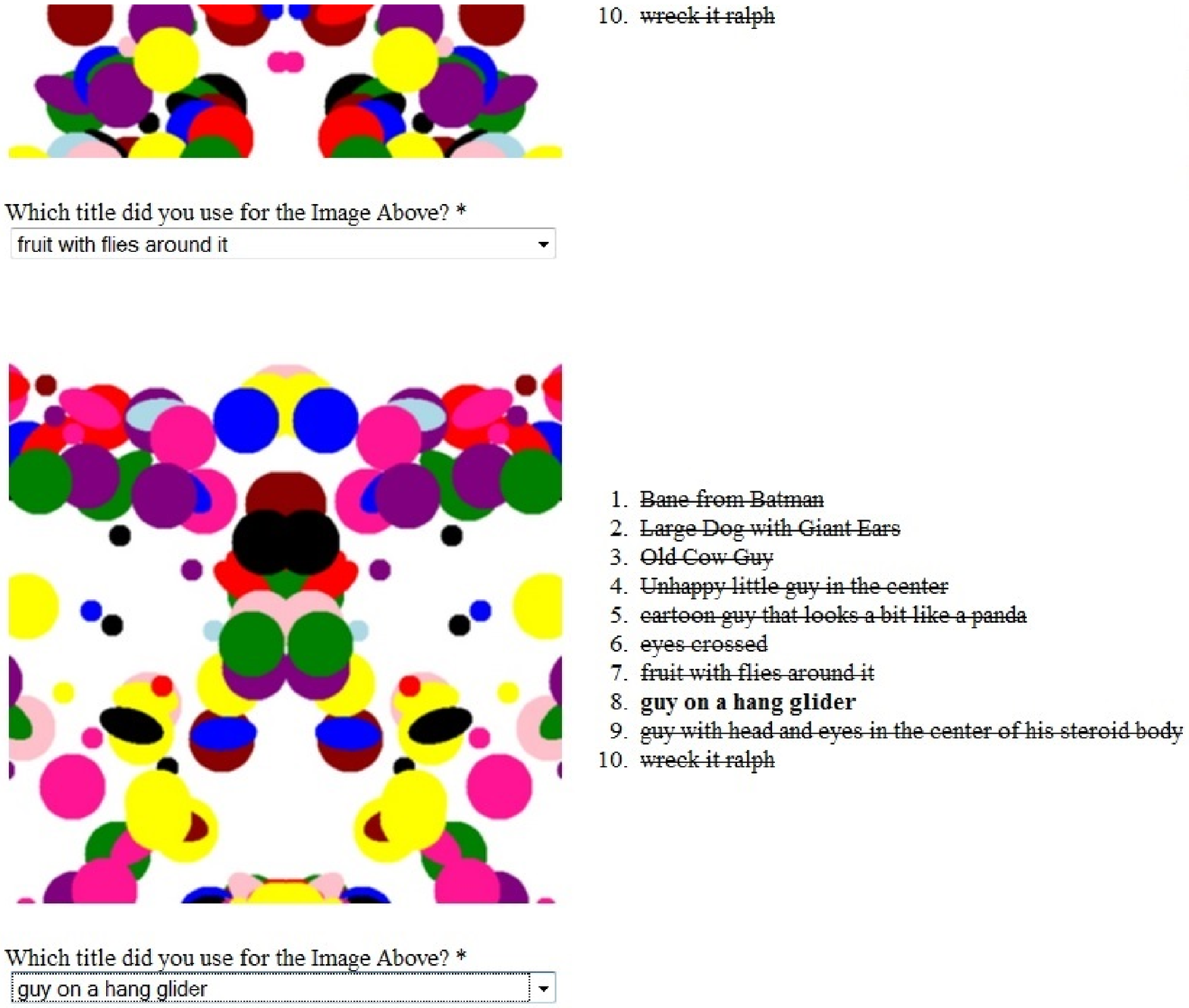}
\caption{Phase 2} \label{fig:userStudyphase2}
\end{figure}

After our participants completed the first phase we waited ten days before asking our participants to return and complete phase 2. During phase 2 we showed each participant the Inkblot images they saw in phase 1 (in a random order) as well as the titles that they created during phase 1 (in alphabetical order). Participants were asked to match the labels with the appropriate image. The purpose of the longer waiting time was to make sure that participants had time to forget their images and their labels. \cut{See figure \ref{fig:userStudyphase2} for an example of phase 2.} Participants were paid an additional $\$1$ for completing phase 2 of the user study. At the beginning of the user study we let participants know that they would be paid during phase 2 even if their answers were not correct. We adopted this policy to discourage cheating (e.g., using screen captures from phase 1 to match the images and the labels) and avoid positively biasing our results. 

  We measured the time it took each participant to complete phase 1. Our results are summarized in table \ref{tab:usabilityPhase1Time}. It is quite likely that some participants left their computer in the middle of the study and returned later to complete the study (e.g., one user took 57.5 minutes to complete the study). While we could not measure time away from the computer, we believe that it is likely that at least $9$ of our participants left the computer. Restricting our attention to the other $61$ participants who took at most 20 minutes we get an adjusted average completion time of $6.2$ minutes.

\begin{table}[t]
\centering
\begin{tabular}{ | c | c | c | }
\hline
 & Phase 1 & Phase 2   \\
 \hline
 Average & 9.3 & 4.5  \\
 \hline
 StdDev  & 9.6  & 3 \\
 \hline
  Max & 57.5 & 18.5  \\
 \hline
  Min & 1.4  & 1.6 \\
\hline
Average $\leq 20$ & 6.2 & N/A \\
   \hline
\end{tabular}
\caption{Completion Times}
\label{tab:usabilityPhase1Time}
\end{table}

Fifty-eight of our participants returned to complete phase 2 by taking our matching test. It took these participants $4.5$ minutes on average to complete the matching test. Seventeen of our participants correctly matched all ten of their labels, and $69\%$ of participants matched at least $5$ out of ten labels correctly. Our results are summarized in table \ref{tab:usabilityResults}.

\begin{table}[t]
\centering
\begin{tabular}{ | p{1in}|p{.7in}| p{1in}| p{1.3in}|}
\hline
$\alpha$-accurate & $\#$ participants & $\frac{\mbox{$\#$ participants}}{58}$ & $\frac{\left| \left\{\pi'~\vline~d_{10}\paren{\pi,\pi'}\leq \alpha \right\}\right|}{10!}$   \\
 \hline
   $\alpha = 0$ & 17 & 0.29 & $2.76 \times 10^{-7}$   \\
 \hline
 $\alpha = 2$ & 22 & 0.38 & $1.27 \times 10^{-5}$   \\
 \hline
 $\alpha = 3 $ & 26 & 0.45 & $7.88 \times 10^{-5}$  \\
\hline
 $\alpha = 4$ & 34 & 0.59 & $6.00 \times 10^{-4}$  \\
   \hline
$\alpha = 5$ & 40 & 0.69 & $3.66 \times 10^{-3}$  \\
\hline
\end{tabular}
\caption{Usability Results: Fraction of Participants who would have authenticated with accuracy parameter $\alpha$}
\label{tab:usabilityResults}
\end{table}

\paragraph{Discussion}  Our user study provides evidence that our construction is at least $\paren{0,0.29}$-usable or $\paren{5,0.69}$-usable. While this means that our Inkblot Matching GOTCHA could be used by a significant fraction of the population to protect their passwords during authentication it also means that the use of our GOTCHA would have to be voluntary so that users who have difficulty won't get locked out of their accounts. Another approach would be to construct different GOTCHAs and allow users to choose which GOTCHA to use during authentication. 

{\bf Study Incentives:} There is evidence that the lack of monetary incentives to perform well on our matching test may have negatively influenced the results (e.g., some participants may have rushed through phase 1 of the study because their payment in round 2 was independent of their ability to match their labels correctly). For example, none of our $18$ fastest participants during phase 1 matched all of their labels correctly, and --- excluding participants we believe left their computer during phase 1 (e.g., took longer than 20 minutes) --- on average participants who failed to match at least five labels correctly took $2$ minutes less time to complete phase 1 than participants who did.

{\bf  Time: } We imagine that some web services may be reluctant to adopt GOTCHAs out of fear driving away customers who don\rq{}t want to spend time labeling Inkblot images \cite{florencio2010security}. However, we believe that for many high security applications (e.g., online banking) the extra security benefits of GOTCHAs will outweigh the costs --- GOTCHAs might even help a bank keep its customers by providing extra assurance that users' passwords are secure. We are looking at modifying our Inkblot generation algorithm to produce Inkblots which require less ``mental effort" to label. In particular could techniques like Perlin Noise \cite{perlin2004implementing} be used to generate Inkblots that can be labeled more quickly and matched more accurately?

{\bf Accuracy:} We believe that the usability of our Inkblot Matching GOTCHA construction can still be improved. One simple way to improve the usability of our GOTCHA construction would be to allow the user to reject Inkblot images that were confusing. We also believe that usability could be improved by providing users with specific strategies for creating their labels (e.g., we found that simple labels like ``a voodoo mask" were often mismatched, while more elaborate stories like ``A happy guy on the ground, protecting himself from ticklers" were rarely mismatched). 

\subsection{An Open Challenge to the AI Community} \label{subsec:OpenChallenge}
We envision a rich interaction between the security community and the artificial intelligence community. To facilitate this interaction we present an open challenge  to break our GOTCHA scheme. 

\paragraph{Challenge Setup} We chose several random passwords \\$\paren{pw_1,...,pw_4} \stackrel{\$}{\leftarrow} \{0, 10^7\}$ and $pw_5 \stackrel{\$}{\leftarrow} \{0, 10^8\}$. We used a function $\mathbf{GenerateInkblots}\paren{pw_i, 10}$ to generate ten inkblots $I^i_{1},...,I^{i}_{10}$ for each password, and we had a human label each inkblot image $\langle\ell_1^i,\ldots,\ell_{10}^i  \rangle\gets H\paren{\langle I^i_{1},\ldots,I^{i}_{10}\rangle,\sigma_0} $. We selected a random permutation $\pi_i:[10]\rightarrow[10]$ for each account, and generated the tuple

\[T_i= \left( s_i, h\paren{pw_i,s_i, \pi_i(1),...,\pi_i(10) },\ell^i_{\pi_i(1)},...,\ell^i_{\pi_i(10)} \right) \ , \]
where $s_i$ is a randomly selected salt value and $h$ is a cryptographic hash function. We are releasing the source code that we used to generate the Inkblots and evaluate the hash function $h$ along with the tuples $T_1,...,T_5$ ---  
see \\ \url{http://www.cs.cmu.edu/~jblocki/GOTCHA-Challenge.html}.

{\noindent \bf Challenge: } Recover each password $pw_i$. 

\paragraph{Approaches} One way to accomplish this goal would be to enumerate over every possible password guess $pw_i'$  and evaluate $h\paren{pw_i',s_i, \pi(1),...,\pi(10) } $ for every possible permutation  $\pi:[10]\rightarrow[10]$. However, the goal of this challenge is to see if AI techniques can be applied to attack our GOTCHA construction. We intentionally selected our passwords from a smaller space to make the challenge more tractable for AI based attacks, but to discourage participants from trying to brute force over all password/permutation pairs we used BCRYPT (Level 15)\footnote{The level parameter specifies the computation complexity of hashing. The amount of work necessary to evaluate the BCRYPT hash function increases exponentially with the level so in our case the work increases by a factor of $2^{15}$.} --- an expensive hash function --- to encrypt the passwords. Our implementation allows the Inkblot images to be generated very quickly from a password guess pw' so an AI program that can use the labels in the password file to distinguish between the correct Inkblots returned by $\mathbf{GenerateInkblots}\paren{pw_i, 10}$ and incorrect Inkblots returned by $\mathbf{GenerateInkblots}\paren{pw_i', 10}$ would be able to quickly dismiss incorrect guesses. Similarly, an AI program which generates a small set of likely permutations for each password guess could allow an attacker to quickly dismiss incorrect guesses. 

\section{Analysis: Cost of Offline Attacks} \label{sec:analysis}
 \label{subsec:offlineAttack}
In this section we argue that our password scheme (protocols \ref{protocol:authenticate} and \ref{protocol:createpwd}) significantly mitigates the threat of offline attacks. An informal interpretation of our main technical result --- Theorem \ref{thm:BruteForce} --- is that either (1) the adversary's offline attack is prohibitively expensive (2) there is a good chance that adversary's offline attack will fail, or (3) the underlying GOTCHA construction can be broken. Observe that the security guarantees are still meaningful even if the security parameters $\epsilon$ and $\delta$ are not negligably small.

\begin{theorem}  \label{thm:BruteForce}
Suppose that our user selects his password uniformly at random from a set $D$ (e.g., $pw \stackrel{\$}{\gets} D$) and creates his account using protocol \ref{protocol:createpwd}. If algorithms \ref{alg:GenerateInkblot} and \ref{alg:GenerateMatchingChallenge} are an $\paren{\epsilon,\delta,\mu}$-GOTCHA then no conservative offline adversary is $\paren{C,\gamma+\epsilon+\delta + \frac{n_H}{|D|},D}$-successful for $C<\gamma|D| 2^{\mu(k)}c_h + n_Hc_H$
\end{theorem}

\begin{proofof}{Theorem \ref{thm:BruteForce}} (Sketch)
We use a hybrid argument.  An adversary who breaches the server is able to recover the tuple $t=\paren{u,r',s, h \paren{u,s,pw,\pi(1),\ldots,\pi(k)},\ell_{\pi(1)},\ldots,\ell_{\pi(k)} } $ as well as the code for the cryptographic hash function $h$ and the code for our GOTCHA --- $\paren{G_1,G_2}$.  

\begin{enumerate}
\item World 0: $W_0$ denotes the real world in which the adversary has recovered the tuple \[ t_0=\paren{u,r',s, h\paren{u,s,pw,\pi(1),\ldots,\pi(k)},\ell_{\pi(1)},\ldots,\ell_{\pi(k)} }\] as well as the code for the cryptographic hash function $h$ and the code for our GOTCHA --- $\paren{G_1,G_2}$. Because the adversary $\mathbf{Adv}$ is conservative it constructs the function

 \[ \mathbf{VerifyHash}\paren{pw',\pi'} = \left\{
\begin{aligned} 
1 && \mbox{if $pw'=pw$ and $\pi'=\pi$} \\
0 && \mbox{otherwise.}
\end{aligned} \ , 
\right.  \]
and uses $\mathbf{VerifyHash}$ as a blackbox. We say that $\mathbf{Adv}$ queries a human $H$ about password $pw'$ if it queries $H$ for $H\paren{\mathbf{GenerateInkblotImages}\paren{1^k,  \mathbf{Extract}\paren{pw',r'}}}$, and we let $D' \subseteq D$ denote the set of passwords for which the adversary queries a human.

\item World 1: $W_1$ denotes a hypothetical world that is similar to $W_0$ except that $\mathbf{VerifyHash}$ function the adversary uses as a blackbox is replaced with the following incorrect version
\begin{multline*} \mathbf{VerifyHash}^1\paren{pw',\pi'} = \\ \left\{
\begin{aligned} 
1 && \mbox{if $pw' \notin D' ,pw'=pw$ and $\pi'=\pi$} \\
0 && \mbox{otherwise.}
\end{aligned} 
\right. \ , 
\end{multline*}

where $D'\subseteq D$ is a subset of passwords which denotes the set of passwords for which the adversary makes queries to a human in the real world.

\item  World 2: $W_2$ denotes a hypothetical world that is similar to $W_1$ except that $\mathbf{VerifyHash}^1$ function the adversary uses as a blackbox is replaced with the following incorrect version
\begin{multline*} \mathbf{VerifyHash}^2\paren{pw',\pi'} = \\ \left\{
\begin{aligned} 
1 && \mbox{if  $\pi'=R\paren{G_2\paren{1^k,\mathbf{Extract}\paren{pw',r'},\ell_1,\ldots,\ell_k}}$ } \\
&& \mbox{and $pw' \notin D' ,pw'=pw$} \\
0 && \mbox{otherwise.}
\end{aligned} 
\right. \ , 
\end{multline*}

where $R$ is a distribution with minimum entropy $\mu(k)$ as in definition \ref{def:GOTCHA}.
\item World 3: $W_3$ denotes a hypothetical real world which is similar to world 2, except that the labels $\ell_{\pi(1)},\ldots,\ell_{\pi(k)}$ are replaced with the labels $\ell_{\pi'(1)}',\ldots,\ell_{\pi'(k)}'$, where $\pi':[k]\rightarrow[k]$ is a new random permutation, and the labels $\ell_i'$ are for a completely unrelated set of Inkblot challenges
\[\ell_1',\ldots,\ell_k \gets H\paren{G_1\paren{1^k,x_1,x_2}} \ , \]
where $x_1,x_2\in \{0,1\}^n$ are freshly chosen random value.
 
\end{enumerate}
In world 3 it is easy to bound the adversary's probability of success. No adversary is $\paren{C,\gamma,D}$-successful for $C <\gamma|D| 2^{\mu(k)}c_h$, because the fake Inkblot labels are not correlated with the actual Inblots that were generated with the real password. Our particular advesary cannot be  $\paren{C,\gamma,D}$-successful for $C <\gamma|D| 2^{\mu(k)}c_h + |D'|c_H$. In
world 2 the adversary might improve his chances of success by looking at the Inblot labels, but by definition of $(\alpha,\beta,\epsilon,\delta,\mu)$-GOTCHA his chances change by at most $\delta$. In world 1 the adversary might further improve his chances of success, but   by definition of $(\alpha,\beta,\epsilon,\delta,\mu)$-GOTCHA his chances improve by at most $\epsilon$. Finally, in world 0 the adversary improves his chances by at most $|D'|/|D|$ by querying the human about passwords in $D'$. 
\end{proofof}

\section{Discussion}
\label{sec:disc}

We conclude by discussing some key directions for future work.

\paragraph{Other GOTCHA Constructions} Because GOTCHAs allow for human feedback during puzzle generation --- unlike HOSPs \cite{canetti2006mitigating} --- our definition potentially opens up a much wider space of potential GOTCHA constructions. One idea might be to have a user rate/rank random items (e.g., movies, activities, foods). By allowing human feedback we could allow the user to dismiss potentially confusing items (e.g., movies he hasn't seen, foods about which he has no strong opinion). There is some evidence that this approach could provide security (e.g., Narayanan and Shmatikov showed that a Netflix user can often be uniquely identified from a few movie ratings \cite{narayanan2008robust}.).

\paragraph{Obfuscating CAPTCHAs} If it were possible to efficiently obfuscate programs then it would be easy to construct GOTCHAs from CAPTCHAs (e.g., just obfuscate a program that returns the CAPTCHA without the answer). Unfortunately, there is no general program obsfuscator \cite{barak2001possibility}. However, the approach may not be entirely hopeless. Point functions \cite{wee2005obfuscating} can be obfuscated, and our application is similar to a point function --- the puzzle generator $G_2$ in an GOTCHA only needs to generate a human solvable puzzle for one input. Recently, multilinear maps have been used to obfuscate conjunctions \cite{obfuscatingConjunctions}  and to obfuscate $NC^1$ circuits \cite{obfuscateIndistinguishability} \footnote{The later result used a weaker notion of obfuscation known as \lq\lq{}indistinguishability obfuscation,\rq\rq{} which (loosely) only guarantees that the adversary cannot distinguish between the obfuscations of two circuits which compute the same function.}. Could similar techniques be used obfuscate CAPTCHAs?

\paragraph{Exploiting The Power of Interaction} Can interaction be exploited and used to improve security or usability in human-authentication?  While interaction is an incredibly powerful tool in computer security (e.g., nonces \cite{rogaway2004nonce}, zero-knowledge proofs \cite{goldreich1999can}, secure multiparty computation \cite{yao1982protocols}) and in complexity theory\footnote{A polynomial time verifier can verify {\bf PSPACE}-complete languages by interacting with a powerful prover \cite{shamir1992ip}, by contrast the same verifier can only check proofs of {\bf NP}-Complete languages without interaction.}, human authentication typically does not exploit interaction with the human (e.g., the user simply enters his password). We view the idea behind HOSPs and GOTCHAs --- exploiting interaction to mitigate the threat of offline attacks --- as a positive step in this direction. %Interaction --- combined with implicit memory ---  has also been used to design a human-authentication scheme that is secure against rubber-hose attacks \cite{rubberHose}. 
	Could interaction be exploited to reduce memory burden on the user by allowing a user to reuse the same secret to authenticate to multiple different servers? The human-authentication protocol of Hopper, et al. \cite{hopper2001secure} --- based on the noisy parity problem --- could be used by a human to repeatedly authenticate over an insecure channel. Unfortunately, the protocol is slow and tedious for a human to execute, and it can be broken if the adversary is able to ask adaptive parity queries \cite{kushilevitz1993learning}.

\bibliography{passwords}

\newpage

\appendix

\section{Missing Proofs} \label{apdx:missingproof}

\begin{reminderclaim}{\ref{claim:correct}}
If $\paren{G_1,G_2}$ is a $\paren{\alpha,\beta,\epsilon,\delta,\mu}$-GOTCHA then at least $\beta$-fraction of humans can sucessfully authenticate using protocol \ref{protocol:authenticate} after creating an account using protocol \ref{protocol:createpwd}.
\end{reminderclaim}

\begin{proofof}{Claim \ref{claim:correct}}
 A legitimate user $H \in \mathcal{H}$ will use the same passwords in protocols  \ref{protocol:createpwd} and \ref{protocol:authenticate}. Hence, 
\[ r_1' = \mathbf{Extract}\paren{pw',r'} = \mathbf{Extract}\paren{pw,r'} = r_1 \ , \] 
and the final matching challenge $\hat{c}_\pi$ is the same one that would be generated by $G_2\paren{1^k,r_1,H\paren{G_1\paren{1^k,r_1,r_2},\sigma_0}}$. If $\hat{c}_\pi$ is consistently solvable with accuracy $\alpha$ by $H$ --- by definition \ref{def:GOTCHA} this is the case for at least $\beta$-fraction of users --- then it follows that
\[d_k\paren{\pi, \pi',\sigma_t} \leq \alpha  \ , \]
where $H\paren{G_1\paren{1^k,r_1,r_2}}$.
For some $\pi_0$ (namely $\pi_0 = \pi$) s.t. $d_k\paren{\pi_0,\pi'}\leq \alpha$ it must be the case that
\begin{eqnarray*}
h_{pw,0} &=& h\paren{u,s,pw',\pi_0(1),...,\pi_0(k)} \\ &=& h\paren{u,s,pw,\pi(1),...,\pi(k)} \\ &=& h_pw \ , 
\end{eqnarray*}

and protocol \ref{protocol:authenticate} accepts.
\end{proofof}

\begin{reminderclaim}{Claim \ref{claim:ClosePermuations}}
For all permutations $\pi:[k]\rightarrow[k]$ and $\alpha \geq 0$ 
\[ \left|\left\{\pi'~\vline~d_k\paren{\pi,\pi'}\leq \alpha \right\} \right| \leq 1 + \sum_{i=2}^\alpha {k \choose i}i! \ .  \]
\end{reminderclaim}

\begin{proofof}{\ref{claim:ClosePermuations}}
It suffices to show that ${k \choose j}j! \geq \left|\left\{ \pi'~\vline~d_k\paren{\pi,\pi'}= j \right\}\right|$. We first choose the $j$ unique indices $i_1,\ldots,i_j$ on which $\pi$ and $\pi'$ differ --- there are   ${k \choose j}$ ways to do this. Once we have fixed our indices $i_1,\ldots,i_j$ we define $\pi'\paren{k} = \pi\paren{k}$ for each  $k \notin \{i_1,\ldots,i_j\}$. Now $j!$ upperbounds the number of ways of selecting the remaining values $\pi'\paren{i_k}$  s.t. $\pi\paren{i_k} \neq \pi'\paren{i_k}$ for all $k \leq j$.
\end{proofof}

\section{HOSP: Pre-Generated CAPTCHAs} \label{apdx:Economics}
The HOSP construction proposed by \cite{canetti2006mitigating} was to simply  fill several high capacity hard drives with randomly generated CAPTCHAs --- discarding the solutions. Once we have compiled a database large $D$ of CAPTCHAs we can use algorithm \ref{alg:G} as our challenge generator --- simply return a random CAPTCHA from $D$. The advantage of this approach is that we can make use of already tested CAPTCHA solutions so there is no need to make hardness assumptions about new AI problems. The primary disadvantage of this approach is that the size of the database $D$ will be limited by economic considerations --- storage isn't free. While $|D|$ the number of CAPTCHAs that could be stored on a hard drive may be large, it is not exponentially large. An adversary could theoretically pay humans to solve every puzzle in $D$ at which point the scheme would be completely broken. 

\begin{algorithm}
\caption{$\mathbf{GenerateChallenge}$}
\begin{algorithmic}
\State {\bf Input:} Random bits $r\in \{0,1\}^n$, Database $D = \{ P_1,...,P_{2^n}\}$ of CAPTCHAs
\State \Return $P_r$
\end{algorithmic}
\label{alg:G}
\end{algorithm}

\paragraph{Economic Cost} Suppose that two 4 TB hard drives are filled will text CAPTCHAS \footnote{At the time of submission a 4 TB hard drive can be purchased on Amazon for less than $\$162$.}. Let $S$ be the space required to store one CAPTCHA, and let $C_H$ denote the cost of paying a human to solve a CAPTCHA. We use the values $S= 8$ KB \footnote{The exact value of $S$ may vary slightly depending on the particular method used to generate the CAPTCHA. When we compressed a text CAPTCHA using popular GIF format the resulting files were consistently $8$ KB.} and $C_H = \$0.001$ \footnote{Motoyama, et al. estimated that spammers paid humans $\$1$ to solve a thousand CAPTCHAs \cite{motoyama2010re}}. In this case $|D| = \frac{4~TB}{8 KB} \approx 10^9$ so we can store a billion unsolved CAPTCHAs on the hard drives. It would cost the adversary $|D| C_H = \$1,000,000$ to solve all of the CAPTCHAs --- or $\$500,000$ to solve half of them. The up front cost of this attack may be large, but once the adversary has solved the CAPTCHAs he can execute offline dictionary attacks against every user who had an account on the server. Many server breaches have resulted in the release of password records for millions of accounts \cite{breach:Zappos,breach:linkedin,breach:sony,breach:rockyou}. If each cracked password is worth between $\$4$ and $\$30$ \cite{passwordBlackMarket} then it may be easily worth the cost to pay humans to solve every CAPTCHA in $D$.

\end{document}